\documentstyle[art10]{article}

\catcode`\@=11
\def\@aabuffer{}
\def\author #1{\expandafter\def\expandafter\@aabuffer\expandafter
{\@aabuffer \small\rm      #1\relax \par}}
\def\address#1{\expandafter\def\expandafter\@aabuffer\expandafter
{\@aabuffer \small\it #1\relax \par\vspace{1em}}}

\def\maketitle{
\begin{center}
   {\bf \@title \par}        
   \vskip 2em                      
   \@aabuffer\relax
\end{center} \par
\gdef\@aabuffer{}
}

\def\abstracts#1{
\begin{center}
{\begin{minipage}{4.5truein}
                 \footnotesize
                 \parindent=0pt #1\par
                 \end{minipage}}\end{center}
                 \vskip 2em \par}

\fussy
\def\section{\@startsection {section}{1}{\z@}{-3.5ex plus -1ex minus 
    -.2ex}{2.3ex plus .2ex}{\bf }}
\def\subsection{\@startsection{subsection}{2}{\z@}{-3.25ex plus -1ex minus 
   -.2ex}{1.5ex plus .2ex}{\it }}

\def\@makefnmark{{$\!^{\@thefnmark}$}}

\renewenvironment{thebibliography}[1]
	{\begin{list}{\arabic{enumi}.}
	{\usecounter{enumi}\setlength{\parsep}{0pt}
	 \setlength{\itemsep}{0pt} \settowidth
	{\labelwidth}{#1.}\sloppy}}{\end{list}}

\newcounter{arabiclistc}

\def\@citex[#1]#2{\if@filesw\immediate\write\@auxout
	{\string\citation{#2}}\fi
\def\@citea{}\@cite{\@for\@citeb:=#2\do
	{\@citea\def\@citea{,}\@ifundefined
	{b@\@citeb}{{\bf ?}\@warning
	{Citation `\@citeb' on page \thepage \space undefined}}
	{\csname b@\@citeb\endcsname}}}{#1}}

\newif\if@cghi
\def\cite{\@cghitrue\@ifnextchar [{\@tempswatrue
	\@citex}{\@tempswafalse\@citex[]}}
\def\citelow{\@cghifalse\@ifnextchar [{\@tempswatrue
	\@citex}{\@tempswafalse\@citex[]}}
\def\@cite#1#2{{$\!^{#1}$\if@tempswa\typeout
	{IJCGA warning: optional citation argument 
	ignored: `#2'} \fi}}

\def\baselinestretch{1.0}
\ifx\selectfont\undefined
\@normalsize\else\let\glb@currsize=\relax\selectfont
\fi

\ifx\selectfont\undefined
\def\@singlespacing{%
\def\baselinestretch{1}\ifx\@currsize\normalsize\@normalsize\else\@currsize\fi%
}
\else
\def\@singlespacing{\def\baselinestretch{1}\let\glb@currsize=\relax\selectfont}
\fi

\long\def\@makecaption#1#2{
   \vskip 10pt 
   \setbox\@tempboxa\hbox{\footnotesize #1: #2}
   \ifdim \wd\@tempboxa >\hsize   
       \leftskip 0pt plus 1fil 
       \rightskip 0pt plus -1fil 
       \parfillskip 0pt plus 2fil 
       \footnotesize #1: #2\par   
     \else                        
       \hbox to\hsize{\hfil\box\@tempboxa\hfil}  
   \fi}

\def\Journal#1#2#3#4{{#1} {\bf #2}, #3 (#4)}

\def\PRD{{\em Phys. Rev.} D}

\def\be{\begin{equation}}
\def\ee{\end{equation}}
\def\bea{\begin{eqnarray}}
\def\eea{\end{eqnarray}}

\begin{document}

\title{GRAVITATIONAL EXCITONS FROM
EXTRA DIMENSIONS}

\author{ U. G\"UNTHER, A. ZHUK }

\address{Department of Physics, University of Odessa,
2 Petra Velikogo Street, Odessa 270100,\\ UKRAINE}




\maketitle\abstracts{
We study inhomogeneous multidimensional cosmological models with a higher
dimensional space-time manifold $M = M_0\times\prod\nolimits_{i=1}
^nM_i$  $( n \ge 1 )$  under dimensional reduction
to $D_0$-dimensional effective models and show that
small inhomogeneous excitations of the scale factors
of the internal spaces near minima of effective potentials 
should be observable as  massive scalar particles (gravitational excitons)
in the external space-time.
}

\mbox{}

\vspace{-6ex}

\section*{Gravitational excitons}

We consider a multidimensional space-time manifold
\begin{equation}
\label{2.1}M = M_0 \times M_1 \times \dots \times M_n  
\end{equation}
with metric
\begin{equation}
\label{2.2}g = g_{MN}(X) dX^M \otimes dX^M =
g^{(0)}+\sum_{i=1}^ne^{2\beta ^i(x)}g^{(i)},
\end{equation}
where $x$ are some coordinates of the $D_0 = d_0+1$ - dimensional
manifold $M_0$.
Let manifolds $M_i$ be $d_i$-dimensional Einstein spaces with metric
$g^{(i)} $, i.e., $R\left[ g^{(i)}\right] =\lambda ^id_i\equiv R_i $.
Internal spaces $M_i \quad (i=1,\dots ,n) $ may have nontrivial
global topology, being compact (i.e. closed and bounded) for any
sign of spatial topology.

With total dimension $D=1+\sum_{i=0}^nd_i$, $\kappa ^2$ a $D$-dimensional
gravitational constant, $\Lambda $ - a $D$-dimensional bare cosmological
constant and $S_{YGH}$ the standard York-Gibbons-Hawking boundary term,
we consider an action of the form
\begin{equation}
\label{2.6}S=\frac 1{2\kappa ^2}\int\limits_Md^DX\sqrt{|g|}\left\{
R[g]-2\Lambda \right\} +S_{add}+S_{YGH}.
\end{equation}
The additional potential term
\begin{equation}
\label{2.7}S_{add}=-\int\limits_Md^DX\sqrt{|g|}\rho (x)
\end{equation}
is not specified and left in its general form, taking into account the
Casimir effect,
the Freund-Rubin monopole ansatz,
or a perfect fluid.
In all these cases $\rho $ depends on the external coordinates through the
scale factors $a_i(x)=e^{\beta ^i(x)}\ (i=1,\ldots ,n)$ of the internal
spaces.

After dimensional reduction the action reads
$$
S=\frac 1{2\kappa _0^2}\int\limits_{M_0}d^{D_0}x\sqrt{|g^{(0)}|}%
\prod_{i=1}^ne^{d_i\beta ^i}\left\{ R\left[ g^{(0)}\right] -G_{ij}g^{(0)\mu
\nu }\partial _\mu \beta ^i\,\partial _\nu \beta ^j+\right.
$$
\begin{equation}
\label{2.8}+\sum_{i=1}^n\left. R\left[ g^{(i)}\right] e^{-2\beta
^i}-2\Lambda -2\kappa ^2\rho \right\} ,
\end{equation}
where $\kappa _0^2=\kappa ^2/V_I $ 
 and
$ V_I =\prod_{i=1}^nv_i=\prod_{i=1}^n\int\limits_{M_i}d^{d_i}y
\sqrt{|g^{(i)}|}$
are the the $D_0$-dimensional gravitational constant 
and the internal space volume.
$G_{ij}=d_i\delta _{ij}-d_id_j$ \\ $(i,j=1,\ldots ,n)$ defines
the midisuperspace metric.
Action (\ref{2.8}) is
written in the Brans-Dicke frame. Conformal transformation
to the Einstein frame
\begin{equation}
\label{2.10}g_{\mu \nu }^{(0)}=\Omega^2 \tilde g^{(0)}_{\mu \nu }
=exp\left( -\frac 2{D_0-2}
\sum_{i=1}^n d_i\beta^i\right) \tilde g^{(0)}_{\mu \nu }
\end{equation}
yields
\begin{equation}
\label{2.12}S=\frac 1{2\kappa _0^2}\int\limits_{M_0}d^{D_0}x\sqrt{|\tilde
g^{(0)}|}\left\{ \tilde R\left[ \tilde g^{(0)}\right]
-\bar G_{ij}\tilde g^{(0)\mu
\nu }\partial _\mu \beta ^i\,\partial _\nu \beta ^j-2U_{eff}\right\} ,
\end{equation}
where
$\bar G_{ij} =d_i\delta _{ij}+\frac 1{D_0-2}d_id_j,\ (i,j=1,\ldots ,n)$,
and the effective potential reads
\begin{equation}
\label{2.15}U_{eff}={\left( \prod_{i=1}^ne^{d_i\beta ^i}\right) }^{-\frac
2{D_0-2}}\left[ -\frac 12\sum_{i=1}^nR_ie^{-2\beta ^i}+\Lambda +\kappa
^2\rho \right] .
\end{equation}
We recall that $\rho $ depends on the scale factors of the internal spaces: $%
\rho =\rho \left( \beta ^1,\ldots ,\beta ^n\right) $. Thus, we are led to
the action of a self-gravitating $\sigma -$model with flat target space
and self-interaction described by the
potential (\ref{2.15}). It can be easily seen that the problem of
the internal spaces stable compactification is reduced now to the search of
models that provide minima of the effective potential (\ref{2.15}).

The midisuperspace metric (target
space metric) 
by a regular coordinate transformation
$\varphi =Q\beta ,\quad \beta =Q^{-1}\varphi \ $
can be turned into a pure Euclidean form
\begin{equation}
\label{2.20}\bar G_{ij}d\beta ^i\otimes d\beta ^j=
\sigma _{ij}d\varphi ^i\otimes
d\varphi ^j=\sum_{i=1}^nd\varphi ^i\otimes d\varphi ^i\, .
\end{equation}

An appropriate transformation $Q:\
\beta ^i\mapsto \varphi ^j=Q_i^j\beta ^i$ is given e.g. by
\begin{equation}
\label{2.21}
\begin{array}{ll}
\varphi ^1 & =-A\sum_{i=1}^nd_i\beta ^i ,\\
&  \\
\varphi ^i & =\left[ d_{i-1}/\Sigma _{i-1}\Sigma _i\right]
^{1/2}\sum_{j=i}^nd_j(\beta ^j-\beta ^{i-1}) ,\quad i=2,\ldots ,n \, ,
\end{array}
\end{equation}
where $\Sigma _i=\sum_{j=i}^nd_j$,
$A=\pm {\left[ \frac 1{D^{\prime }}\frac{D-2}{D_0-2}\right] }^{1/2}$
and $D^{\prime }=\sum_{i=1}^nd_i$. So we can write action (\ref{2.12}) as
\begin{equation}
\label{2.23}S=\frac 1{2\kappa _0^2}\int\limits_{M_0}d^{D_0}x\sqrt{|\tilde
g^{(0)}|}\left\{ \tilde R\left[ \tilde g^{(0)}\right] -\sigma _{ik}\tilde
g^{(0)\mu \nu }\partial _\mu \varphi ^i\,\partial _\nu \varphi
^k-2U_{eff}\right\}
\end{equation}
with effective potential
\begin{equation}
\label{2.24}U_{eff}=e^{\frac 2{A(D_0-2)}\varphi ^1}\left( -\frac
12\sum_{i=1}^nR_ie^{-2{(Q^{-1})^i}_k\varphi ^k}+\Lambda +\kappa ^2\rho
\right) .
\end{equation}

Let us suppose that this potential has minima which are localized at
points $\vec \varphi_c,c=1,...,m$ : 
$\left. \frac {\partial U_{eff}}{\partial \varphi^i}
\right|_{\vec \varphi_c} = 0$. Then, for small field fluctuations
$\xi^i \equiv \varphi^i - \varphi^i_{(c)}$ around the minima the
potential (\ref{2.24}) reads
\begin{equation}
\label{2.25}U_{eff}=
U_{eff}\left( \vec \varphi _c\right) +\frac 12\sum_{i,k=1}^n\bar
a_{(c)ik}\xi ^i\xi ^k+O(\xi ^i\xi ^k\xi ^l) \, ,
\end{equation}
where the Hessians
$\bar a_{(c)ik}:=\left. \frac{\partial ^2U_{eff}}{\partial \xi
^i\,\partial \xi ^k}\right| _{\vec \varphi _c} $
are assumed as not vanishing identically. 
The action functional (\ref{2.23}) reduces now
to a family of action functionals for fluctuation fields $\xi ^i$%
\begin{equation}
\label{2.27}
\begin{array}{ll}
S= & \frac 1{2\kappa _0^2}\int\limits_{M_0}d^{D_0}x
\sqrt{|\tilde g^{(0)}|}\left\{ \tilde R\left[ \tilde g^{(0)}\right]
-2U_{eff}\left( \vec \varphi _c\right) -\right. \\  &  \\
& \left. -\sigma _{ik}\tilde g^{(0)\mu \nu }\partial _\mu \xi ^i\,\partial
_\nu \xi ^k-\bar a_{(c)ik}\xi ^i\xi ^k\right\} ,\ c=1,...,m.
\end{array}
\end{equation}
It remains to diagonalize the Hessians $\bar a_{(c)ik}$ by appropriate  $%
SO(n)-$ro\-ta\-tions $S_c:\ \xi \mapsto \psi =S_c\xi ,\quad S_c^{\prime
}=S_c^{-1}$%
\begin{equation}
\label{2.28}\bar A_c=S_c^{\prime }M_c^2S_c,\quad M_c^2={\rm diag\ }%
(m_{(c)1}^2,m_{(c)2}^2,\ldots ,m_{(c)n}^2),
\end{equation}
leaving the kinetic term $\sigma _{ik}\tilde g^{(0)\mu \nu }\partial _\mu \xi
^i\,\partial _\nu \xi ^k$ invariant
\begin{equation}
\label{2.29}\sigma _{ik}\tilde g^{(0)\mu \nu }\partial _\mu \xi ^i\,\partial
_\nu \xi ^k=\sigma _{ik}\tilde g^{(0)\mu \nu }\partial _\mu \psi ^i\,\partial
_\nu \psi ^k,
\end{equation}
and we arrive at action functionals for decoupled normal modes of linear $%
\sigma -$models in the background metric $\tilde g^{(0)}$ of the external
\mbox{space-time:}
\begin{eqnarray}\label{2.30}
S & = & \frac{1}{2\kappa _0^2}\int \limits_{M_0}d^{D_0}x \sqrt
{|\tilde g^{(0)}|}\left\{\tilde R\left[\tilde g^{(0)}\right] - 2\Lambda
_{(c)eff}\right\} + \nonumber\\
\ & + & \sum_{i=1}^{n}\frac{1}{2}\int \limits_{M_0}d^{D_0}x \sqrt
{|\tilde g^{(0)}|}\left\{-\tilde g^{(0)\mu \nu}\psi ^i_{,\mu}\psi
^i_{,\nu} -
m_{(c)i}^2\psi ^i\psi ^i\right\},\ c=1,\ldots ,m,
\end{eqnarray}
where $\Lambda _{(c)eff}\equiv U_{eff}\left(
\vec \varphi _c\right) $ is the $D_0$-dimensional effective cosmological
constant and the factor $\sqrt{V_I /\kappa ^2}$ has been
included into $\psi $ for convenience: $\sqrt{V_I /\kappa ^2}\psi
\rightarrow \psi $.

Thus, conformal excitations of the metric of the internal spaces behave as
massive scalar fields developing on the background of the external 
space-time. By analogy with excitons in solid state physics where they are
excitations of the electronic subsystem of a crystal, the excitations of the
internal spaces were  called gravitational excitons \cite{gz}.

\end{document}